\documentclass{aa}
\usepackage{graphicx}
\usepackage{txfonts}
\usepackage{natbib}
\bibpunct{(}{)}{;}{a}{}{,}
\newcommand{\Teff}{\ensuremath{T\mathrm{\hspace*{-0.4ex}_{eff}}}\,}
\newcommand{\sdsslong}{SDSS~J122859.93$+$104032.9}
\newcommand{\sdss}{SDSS\,J1228+1040}

\begin{document}

\title{The gaseous debris disk of the white dwarf \sdss\thanks{Based on observations with the NASA/ESA \emph{Hubble} Space Telescope, obtained at the Space Telescope Science Institute, which is operated by the Association of Universities for Research in Astronomy, Inc., under NASA contract \mbox{NAS5-26666}.}}

\subtitle{HST/COS search for far-ultraviolet signatures}

\author{S. Hartmann \and T. Nagel \and T. Rauch \and K. Werner}

\institute{Institute for Astronomy and Astrophysics, Kepler Center for Astro and Particle Physics, Eberhard Karls University, Sand~1, 72076~T\"ubingen, Germany\\ \email{hartmann@astro.uni-tuebingen.de}}

\date{Received xx xx 2016 / Accepted xx xx 2016}

\abstract
    {Gaseous and dust debris disks around white dwarfs (WDs) are formed from tidally disrupted planetary bodies. This offers an opportunity to determine the composition of exoplanetary material by measuring element abundances in the accreting WD's atmosphere. A more direct way to do this is through spectral analysis of the disks themselves.}
    {Currently, the number of chemical elements detected through disk emission-lines is smaller than that of species detected through lines in the WD atmospheres. We assess the far-ultraviolet (FUV) spectrum of one well-studied object (\sdsslong) to search for disk signatures at wavelengths $<1050$\,\AA, where the broad absorption lines of the Lyman series effectively block the WD photospheric flux. In addition, we investigate the \ion{Ca}{ii} infrared triplet (IRT) line profiles to constrain disk geometry and composition.}
    {We performed FUV observations (950--1240\,\AA) with the \emph{Hubble} Space Telescope/Cosmic Origins Spectrograph and used archival optical spectra. We compared them with \mbox{non-local} thermodynamic equilibrium model spectra.}
    {No disk emission-lines were detected in the FUV spectrum, indicating that the disk effective temperature is $\Teff\approx 5000$\,K. The long-time variability of the \ion{Ca}{ii} IRT was reproduced with a precessing disk model of bulk Earth-like composition, having a surface mass density of $0.3$\,g\,cm$^{-2}$ and an extension from 55 to 90 WD radii. The disk has a spiral shape that precesses with a period of approximately 37 years, confirming previous results.}
    {}


    \maketitle
    %

    \begin{figure*}
      \centering
      \includegraphics[angle=-90,width=\textwidth]{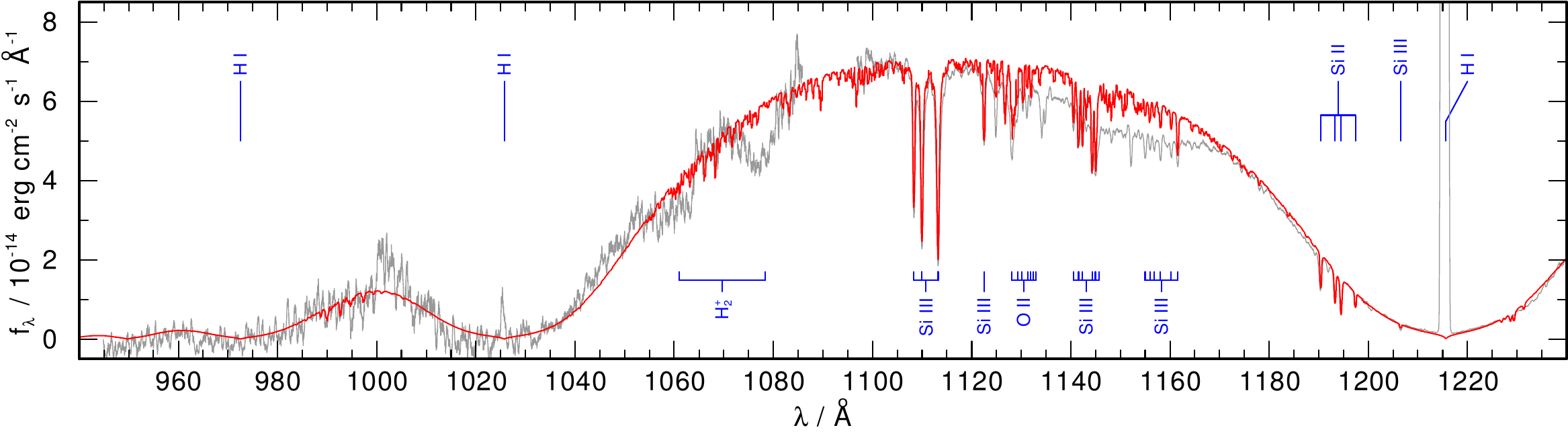}
      \caption{Observation from HST/COS of \sdss\ (thin gray line) and the WD model (thick red line). The spectra are convolved with a 0.4\,\AA\ wide boxcar. Some prominent spectral lines are marked.}
      \label{fig:obs}
    \end{figure*}

    \section{Introduction}

    White dwarfs (WDs) that cool down to effective temperatures $\Teff\approx 25\,000\,$K should have either pure H or pure He atmospheres, as a result of their high surface gravity. Heavy elements sink out of the atmosphere toward the stellar interior. However, a significant fraction (20--30\%) displays photospheric absorption lines from metals \citep[e.g.,][]{Koester:2005}. These polluted WDs must actively accrete matter at a rate of the order of $10^8\,$g\,s$^{-1}$ to sustain the atmospheric metal content, because diffusion time scales are by orders of magnitude shorter than the WD cooling age \citep[e.g.,][]{Koester:2009,Koester:2014}.

    After the discovery of warm dust disks \citep[e.g.,][and references therein]{Farihi:2012} and gaseous disks \citep{Gaensicke:2006,Wilson:2014} around many polluted WDs, it is now commonly accepted that accretion occurs from debris disks that are located within the WD tidal volume. The disks contain material from tidally disrupted exoplanetary bodies that were scattered towards the central star as a consequence of a dynamical resettling of a planetary system in the post-main sequence phase \citep{Debes:2002,Jura:2003}. Therefore, the metal abundance pattern in the polluted WD atmospheres allows to conclude on the composition of the accreted matter. This opened up the exciting possibility to study the composition of extrasolar planetary material. Generally, the abundances are similar to those found in solar system objects \citep{Jura:2014}.

    Concluding on the composition of the accreted material from the WD photospheric abundance pattern is not straightforward. The results are based on the knowledge of metal diffusion rates in WD atmospheres and envelopes \citep{Koester:2009}. Other uncertainties enter, for example the depth of the surface-convection zone. It is therefore desirable to seek alternative methods to determine the chemical composition of the accreted material. Ideally, it is derived by direct observation of the accreting matter.

    One possibility of doing this is offered by line spectroscopy of gaseous disks. Their hallmark is the double-peaked emission lines from the \ion{Ca}{ii} infrared triplet \citep[IRT; $\lambda\lambda$\,8498, 8542, 8662\,\AA,][]{Gaensicke:2006}. Detailed investigations have shown that the gaseous and dusty disks are roughly spatially coincident concerning the radial distance from the WD \citep{Brinkworth:2009,Farihi:2010,Melis:2010}; a result that is consistent with the scenario in which the disk material has its origin in disrupted planetary bodies. The \ion{Ca}{ii} IRT line profiles were scrutinized to derive disk geometries and secular evolution. The most detailed investigation in this respect represents the Doppler imaging of \sdsslong\ (henceforth \sdss) based on spectra taken over twelve years \citep{Manser:2016}.

    The spectral analysis of the gas disks with \mbox{non-local} thermodynamic equilibrium (\mbox{non-LTE}) radiation transfer models aims to probe the physical structure and chemical composition of the disks. It has been shown that the disks have effective temperatures of the order $\Teff=6000$\,K and that they are strongly hydrogen-deficient \citep[H $< 0.01$ mass fraction,][]{Hartmann:2011}. Currently, the prospects of studing the gas disk composition are hampered by the fact that only few species were actually identified. In the best-studied case, \sdss, disk emissions from four elements (O, Mg, Ca, Fe) were detected, whereas a total of eight elements were detected in the WD spectrum \citep{Manser:2016}. Ultraviolet (UV) spectroscopy was expected to reveal more species, but was unsuccessful \citep{Gaensicke:2012}.

    The general problem is that the flux of the relatively hot WDs increases towards the UV. Therefore, we performed observations of \sdss\ in the \mbox{far-UV} (FUV) range (i.e., 950--1230\,\AA), where the WD flux is strongly depressed by broad photospheric hydrogen Lyman lines, striving to discover gas disk signatures in this wavelength region.

    In the following (Sect.\,\ref{sec:obj}) we summarize the current knowledge of \sdss. Section\,\ref{sec:obs} describes the observations used in this work. In Sect.\,\ref{sec:methods} we introduce the spectral modeling of the WD and the gas disk. In Sect.\,\ref{sec:results} we compare the FUV observations with our models. We also use the disk model spectra to fit the \ion{Ca}{ii} IRT in observations performed in 2003 and 2014. We summarize our results in Sect.\,\ref{sec:summary}.

    \section{White dwarf \sdss}\label{sec:obj}

    The \mbox{DAZ-type} WD \sdss\ was the first WD discovered that was surrounded by a gaseous metal disk \citep{Gaensicke:2006}. The double-peaked line profiles of the IRT have a \mbox{peak-to-peak} separation that indicates a Keplerian rotation velocity of $v\,\sin\,i=315\,\mathrm{km\,s}^{-1}$ (with $i\approx 70^{\circ}$). Two other weak emission features of \ion{Fe}{ii} $\lambda\lambda$\,5018/5169\,\AA\ were seen by \citet{Gaensicke:2006}. Weak emissions from the \ion{Ca}{ii} H \& K lines were discovered by \citet{Melis:2010}. The \ion{Mg}{ii} emission line (resonance doublet $\lambda\lambda$\,2796/2804\,\AA) was detected in \emph{Hubble} Space Telescope/Cosmic Origin Spectrograph (HST/COS) data \citep{Hartmann:2011,Manser:2016}. Recently, \citet{Manser:2016} found additional emission lines (\ion{Ca}{ii}, \ion{O}{i}, \ion{Mg}{i--ii}, \ion{Fe}{ii}) in deep coadded optical spectra, increasing the total number of observed gaseous elements in this system to four.

    The line profiles of the \ion{Ca}{ii} IRT exhibit an asymmetry \citep{Gaensicke:2006} that is variable in time \citep{Melis:2010}, indicating a \mbox{non-axisymmetric} distribution of the line-emitting region. \mbox{Non-LTE} modeling of vertical disk structure and emergent spectra of spiral-arm like distributions suggest a disk extension from 58--135 WD radii ($R_{\mathrm{WD}}$), a surface mass density of $\varSigma\approx 0.3$\,g\,cm$^{-2}$ and $\Teff\approx 6000$\,K \citep{Hartmann:2011}. The disk extension is roughly consistent with the inner and outer disk radii determined by \citet{Gaensicke:2008} and \citet{Melis:2010}, respectively, namely 40\,$R_{\mathrm{WD}}$ and 108\,$R_{\mathrm{WD}}$. A very detailed Doppler imaging analysis performed by \citet{Manser:2016} surveyed the disk evolution over twelve years, showing that the IRT line profile changes can be interpreted as the precession of a fixed emission pattern with a period in the range 24--30\,a. They suggest that the precession is due to general relativistic effects.

    An infrared excess was reported by \citet{Brinkworth:2009}, providing evidence for a cool, metal-rich dust disk. They derive a radial extension of the dust disk from 18\,$R_{\mathrm{WD}}$ to 107\,$R_{\mathrm{WD}}$ and $\Teff=1670$\,K to 450\,K from the inner to the outer disk rim. Hence, gaseous and dust disks have almost the same extension.

    As to the WD itself, \citet{Gaensicke:2012} observed it with HST/COS in the wavelength range 1130--1795\,\AA. They report photospheric absorption lines of C, O, Al, Si, Cr and Ni. As a compromise from fitting optical (Sloan Digital Sky Survey, SDSS) and the UV spectra separately, they derived $\Teff=20\,900\pm 900$\,K, $\log\,(g/\mathrm{cm\,s}^{-2})=8.15\pm 0.04$. The metal abundances derived for the accreted material is closely bulk Earth-like. In particular, the extremely low carbon abundance is in accordance with the fact that the disk was found to be strongly carbon depleted \citep{Hartmann:2011}.

    \section{Observations}\label{sec:obs}

\begin{figure}
      \centering
      \includegraphics[width=0.9\columnwidth]{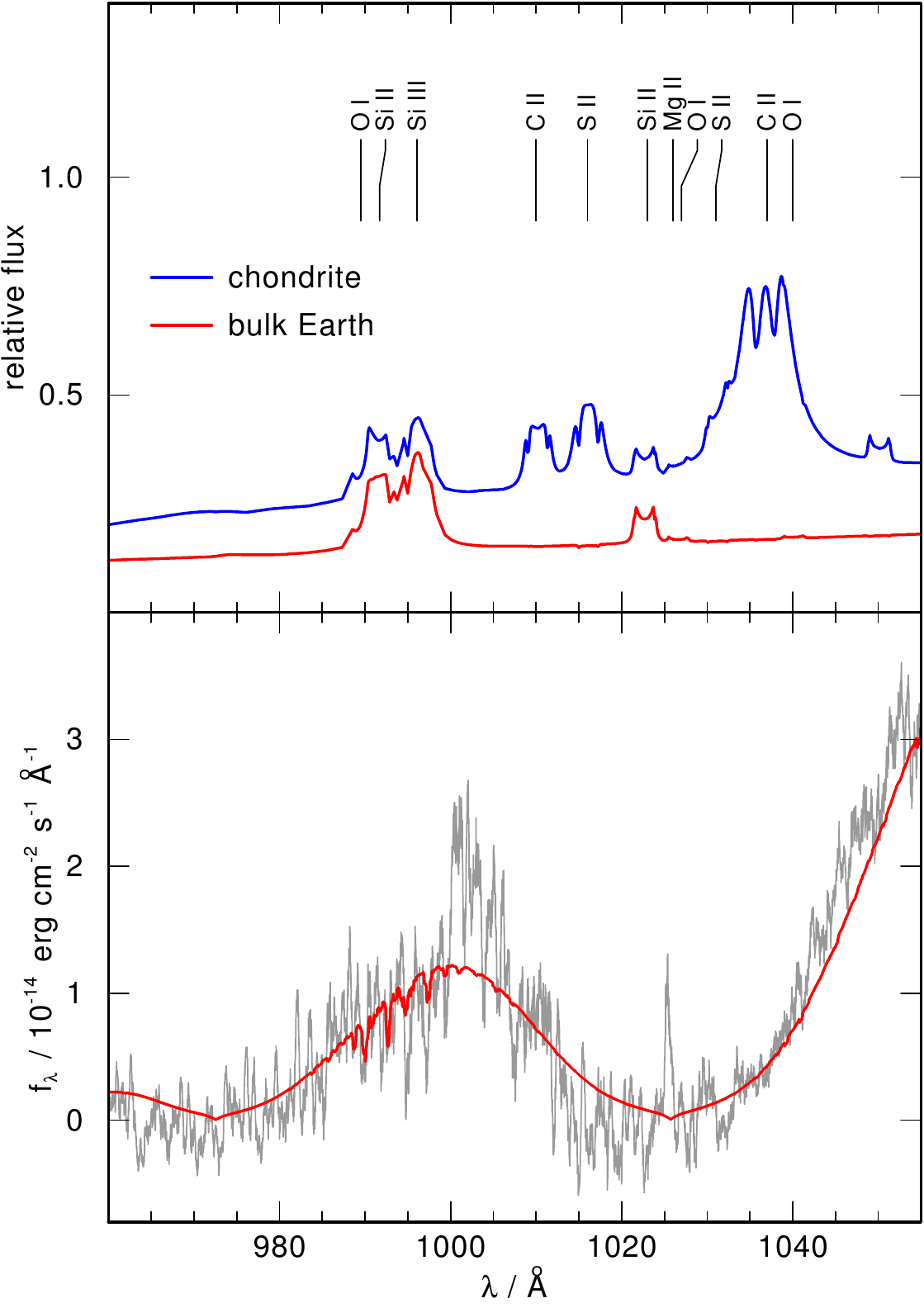}
      \caption{\textit{Bottom panel:} Detail of the observed spectrum of \sdss\ (thin, black line) and a WD model (thick, red line). \textit{Top panel:} Two disk-model spectra with CI chondrite and bulk Earth mixture. Line positions are indicated.}
      \label{fig:obsdisk}
    \end{figure}

    The FUV observation (dataset LBXT51010, total exposure time 10\,093\,s) was performed on June 19, 2013, using HST/COS with grating G130M (central wavelength setting 1096\,\AA), where segments B and A cover 940--1081\,\AA\ and 1096--1237\,\AA, respectively. The spectrum, shifted to rest wavelength to account for the WD's radial velocity plus gravitational redshift of $v=38$\,km\,s$^{-1}$, is shown in Fig.\,\ref{fig:obs} together with a WD model. A detail in the vicinity of the Ly\,$\beta$ and Ly\,$\gamma$ lines is shown in Fig.\,\ref{fig:obsdisk}. In the range 1130--1170\,\AA, the observation has a broad depression which is not exhibited by the model. We have compared this spectrum with an archival spectrum taken with the same setup but a different wavelength setting (dataset LB5Z03010, taken on April~12, 2010, PI: B.~G\"ansicke), such that the blue edge is located at 1135\,\AA\ and hence it overlaps with our spectrum. It does not exhibit the flux depression and is well fit by the WD model. In addition, the absolute flux levels of both datasets differ by about 10\%, (the older spectrum has a lower flux level), pointing at flux calibration problems in our observations. For completeness, we note that our observation was a repetition of a previous attempt that failed because of an observatory error during target acquisition peakup. The resulting spectrum (dataset LBXT01010) is degraded because the target was not centered in the slit.

    We use two archival spectra to apply our disk models to the variable \ion{Ca}{ii} IRT. In March 2003, a spectrum was recorded by the SDSS, and another one was taken in June 2014 with the \mbox{X-shooter} instrument at the Very Large Telescope of the European Southern Observatory \citep{Vernet:2011}.

    \section{Methods}\label{sec:methods}

    \subsection{Disk models}

    For the calculation of the synthetic disk spectra, we used our accretion-disk code AcDc \citep{Nagel:2004}. In the past, it has been successfully applied to the analysis of disks in cataclysmic variables \citep{Kromer:2007}, \mbox{C/O-dominated} disks in ultracompact \mbox{X-ray} binaries \citep{Werner:2006}, and \mbox{Fe-dominated} supernova-fallback disks \citep{Werner:2007}. More recently, we also used it for the analysis of gaseous debris disks around WDs \citep{Hartmann:2011,Hartmann:2014}.

    The disk model assumes radial symmetry and is divided in concentric rings with radius $R$. For each ring we choose values for $\Teff(R)$ and $\varSigma(R)$. These quantities internally determine a mass-accretion rate $\dot{M}$ and a kinematic viscosity $\nu$ in such a way as to fulfill the formulation for a viscous $\alpha$ disk by \citet{Shakura:1973}:
      \begin{align}
        T_{\mathrm{eff}}^{4}(R)&=\frac{3GM_{\mathrm{WD}}\dot{M}}{8\piup R^{3}}\left(1-\sqrt{\frac{R_{\mathrm{WD}}}{R}}\right),\\
        \nu\varSigma(R)&=\frac{\dot{M}}{3\piup}\left(1-\sqrt{\frac{R_{\mathrm{WD}}}{R}}\right)\quad\text{,}
      \end{align}
      where $M_{\mathrm{WD}}$ and $R_{\mathrm{WD}}$ are the mass and radius of the central WD, while $G$ is the gravitational constant.

      Following \citet{Nagel:2004} a set of equations that delivers the vertical structure (temperature, particle densities) and the spectral energy distribution is solved. This set comprises: radiation-transfer equations for several 10\,000 frequencies, hydrostatic equation, energy balance, and several 100 \mbox{non-LTE} rate equations for atomic level populations. The numerical solution of the equations is achieved with an accelerated lambda iteration method \citep{Werner:1986}.

      In the vertical structure calculation we proceed as if the emitted radiation is viscously generated, although the actual disk heating mechanism is unknown. The internally used accretion rate needed to yield the envisaged \Teff\ is much higher than one would expect for the metallic gas disks considered here. One alternative heating mechanism could be energy dissipation through disk asymmetries \citep{Jura:2008}, another one may be that of a \mbox{so-called} \ion{Z}{ii} region, where the metals are photoionized by UV photons emitted from the WD \citep{Melis:2010}.

      For comparison with observations, the emergent spectra are velocity shifted to account for Keplerian motion, hence, $R$ and disk inclination $i$ appear as additional kinematic and dynamic parameters. Finally, the modeled ring fluxes are summed up and corrected for the apparent disk surface area due to the inclination of the disk. At this stage deviations from radial symmetry can be accounted for by representing, for example, spiral arms by patching only ring segments. Such structures follow from hydrodynamical simulations. These simulations also predict spectral line shape variability that can be compared to observations \citep{Hartmann:2011}.

    \begin{figure}
      \centering
      \includegraphics[width=0.9\columnwidth]{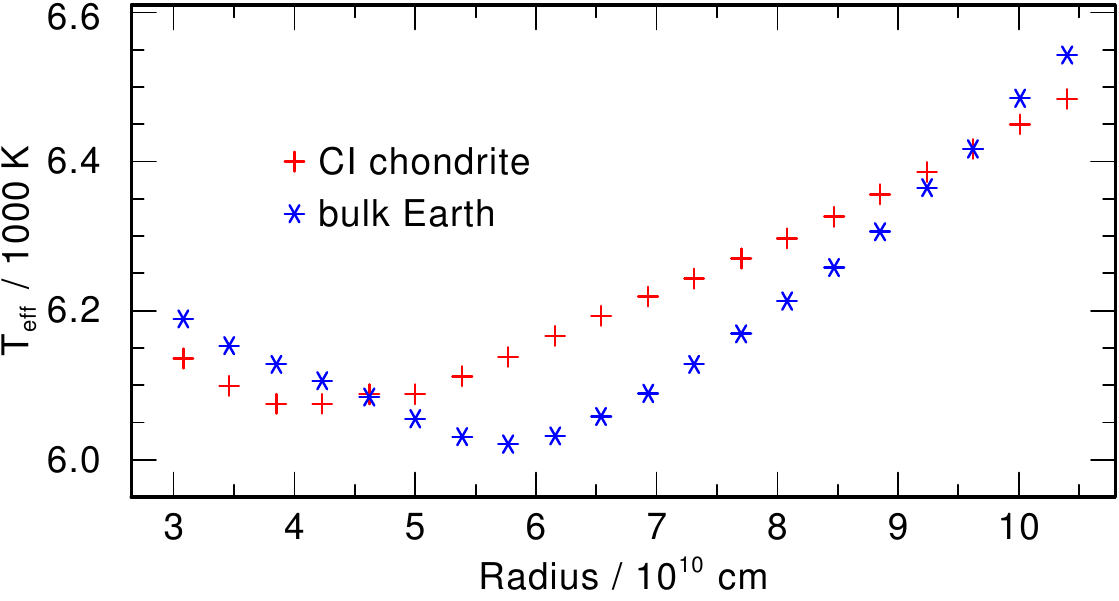}
      \caption{Radii and effective temperatures of the twenty concentric rings of two disk models with different chemical compositions.}
      \label{fig:radius}
    \end{figure}

      For \sdss\ we adopt $R_{\mathrm{WD}}=0.011\,R_{\odot}$ and $M_{\mathrm{WD}}=0.77\,M_{\odot}$. The disk model is composed by twenty rings, extending from $R_{\mathrm{in}}=40\,R_{\mathrm{WD}}$ to an outer radius of $R_{\mathrm{out}}=135\,R_{\mathrm{WD}}$. Building on our previous work \citep{Hartmann:2011}, we went for radial constant values of $\varSigma=0.3$\,g\,cm$^{-2}$ and $\Teff=6000$\,K. Numerical conversions during the procedure may alter the chosen \Teff by a little. The resulting radial run of \Teff is depicted in Fig.\,\ref{fig:radius}.

    The models were calculated for a chemical composition comparable to CI chondrites in the Solar system (H = 10$^{-8}$, C = 0.047, O = 0.601, Mg = 0.126, Si = 0.141, S = 0.072, Ca = 0.024, mass fractions) and for a bulk Earth mixture (H = 10$^{-8}$, C = 10$^{-8}$, O = 0.482, Mg = 0.235, Si = 0.258, S = $8.4\times 10^{-5}$, Ca = 0.025). In the case of the bulk Earth mixture, the sulfur abundance is reduced by a factor of hundred compared to the standard value. This is required to avoid the appearance of strong, unobserved \ion{S}{i} emission lines ($^5$P--$^5$D$^{\textrm{o}} \lambda\lambda$\,8670--8595\,\AA\ multiplet) overlapping with the reddest IRT component \citep{Hartmann:2014}. Atomic data that we used for our \mbox{non-LTE} calculations are provided by model atoms \citep[cf.,][]{Rauch:2003} and are summarized in Table~\ref{tab:modelatom}. They were taken from \emph{TMAD}, the T\"ubingen model atom database\footnote{http://astro.uni-tuebingen.de/\raisebox{.2em}{\tiny $\sim$}TMAD}. In some cases, it was diffucult to converge the single-disk ring models and to overcome numerical instabilities, so particular line transitions had to be removed from the model atom during the iteration procedure. Except for the \ion{Si}{iii} $\lambda$\,1206\,\AA\ line, this measure was restricted to far-infrared lines of \ion{O}{i} and \ion{Si}{ii} with $\lambda > 20\,000$\,\AA.

    \begin{table}
      \begin{center}
        \caption{Number of \mbox{non-LTE} levels and lines of the model ions used in
          our disk models.\tablefootmark{a}}
        \label{tab:modelatom}
        \small
        \begin{tabular}{ccccc}\hline\hline
          \noalign{\smallskip} & I & II & III & IV\\\hline
          \noalign{\smallskip} H &10, 45 & 1, 0\\
          C &15, 19 & 17, 32 & 1, 0\\
          O &14, 13 & 16, 26 & 1, 0\\
          Mg &17, 15 & 15, 29 & 1, 0\\
          Si &19, 29 & 20, 35 & 17, 27 & 1, 0\\
          S &33, 52 & 23, 37 & 1, 0\\
          Ca & 7, 3 & 14, 21& 1, 0 \\
          \noalign{\smallskip} \hline
        \end{tabular}
        \tablefoot{ \tablefoottext{a}{The first and second number of each table entry denote the number of levels and lines per ionisation stage, respectively.} }
      \end{center}
    \end{table}

    \subsection{WD model}

    For the WD, we calculated a plane-parallel \mbox{non-LTE} model atmosphere ($\Teff=20\,900$\,K, $\log\,(g/\mathrm{cm\,s}^{-2})=8.15$) with the T\"ubingen model atmosphere package \citep[\emph{TMAP},][]{Werner:2003,Werner:2012}. Besides hydrogen, it includes the following elements with abundances based on the results of \citet{Gaensicke:2012}: C = $3.8 \times 10^{-7}$, O = $4.5 \times 10^{-4}$, Mg = $1.9 \times 10^{-4}$, Si = $1.8 \times 10^{-4}$, Fe = $3.4 \times 10^{-4}$, Ni = $1.8 \times 10^{-5}$ (mass fractions). A generic model atom considering opacities from all other iron group elements with solar abundance ratio relative to iron was included.

    \section{Results}\label{sec:results}

    \subsection{FUV}

    The observed spectrum and the WD model spectrum are shown in Fig.\,\ref{fig:obs}. The overall shape is well reproduced by the WD spectrum and its broad \ion{H}{i} Lyman lines. Emission cores in Ly\,$\alpha$ and Ly\,$\beta$ are of geocoronal origin. Our model does not include the two broad H$_{2}^{+}$ Lyman satellites that are seen in the observation. There is no indication of any disk emission line.

    As the WD flux in the FUV is strongly depressed by broad photospheric Lyman lines (Ly\,$\beta$ and higher series members), our disk model with $\Teff=6000$\,K and CI chondrite mixture predicts detectable emission lines from carbon and sulfur (\ion{C}{ii} $\lambda$\,1037\,\AA, \ion{C}{ii} $\lambda$\,1010\,\AA, \ion{S}{ii} $\lambda$\,1016\,\AA). The disk model with bulk Earth mixture and reduced S abundance (suggested by the optical spectrum, see above) explains this finding. The \mbox{non-detection} of the predicted silicon lines (\ion{Si}{ii} $\lambda\lambda$\,990/993\,\AA\ and $\lambda\lambda$\,1021/1024\,\AA, \ion{Si}{iii} $\lambda\lambda$\,994/995/997\,\AA), however, calls for another explanation. We have computed a cooler disk model with $\Teff=5000$\,K and found that its FUV flux is significantly lower and all emission lines disappeared. The lower \Teff\ generally explains the paucity of disk emission-lines in the UV, where only the \ion{Mg}{ii} resonance doublet in the \mbox{near-UV} ($\lambda\lambda$\,2796/2804\,\AA) was detected.

    \begin{figure}
      \centering
      \includegraphics[width=0.9\columnwidth]{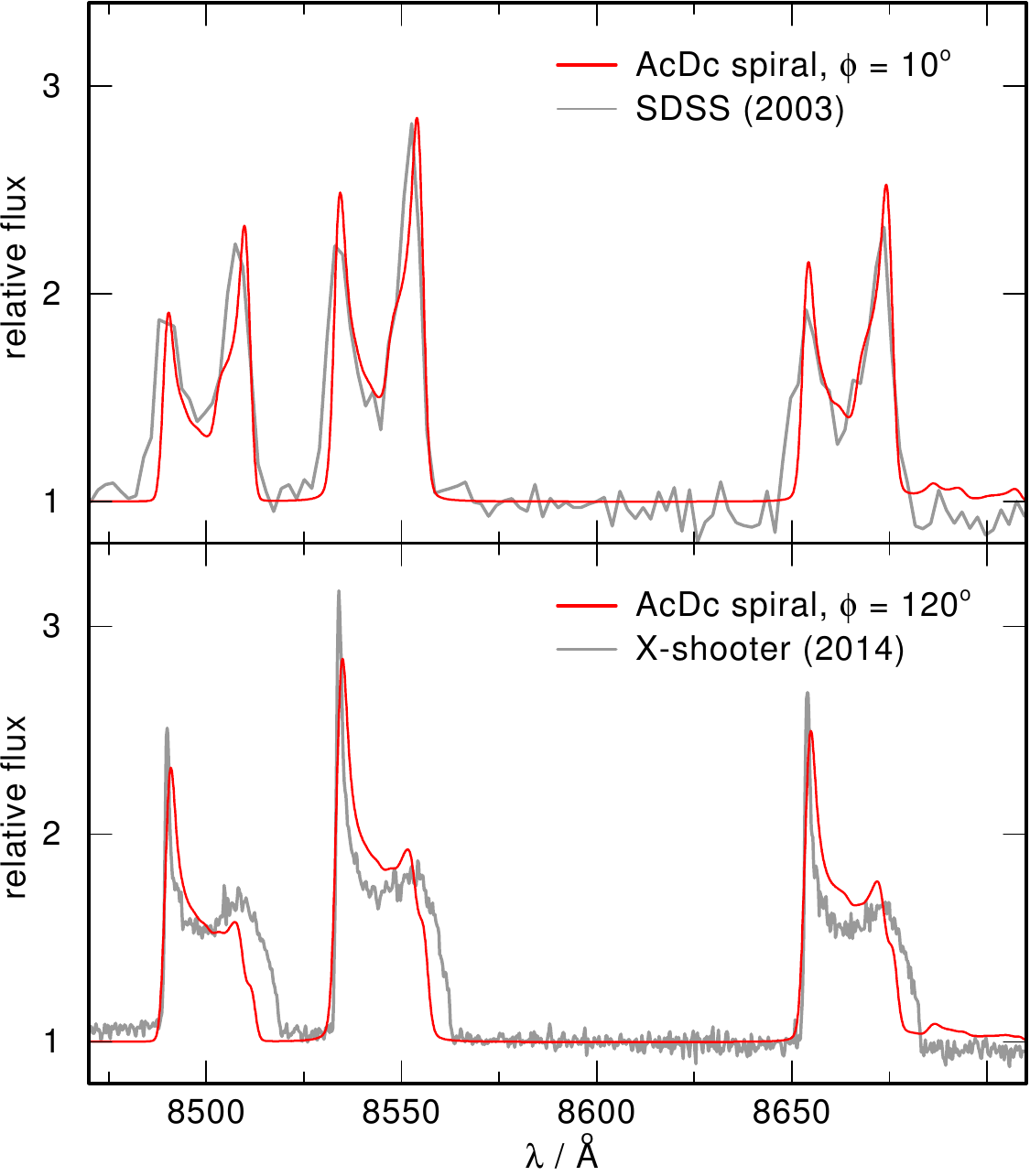}
      \caption{Long-time variability of the \ion{Ca}{ii} infrared triplet in \sdss. \emph{Upper panel:} SDSS spectrum (2003) compared with a spiral-shape accretion-disk model spectrum. \emph{Lower panel}: \mbox{X-shooter} spectrum (2014) compared with the same disk model but observed from a different viewing angle.}
      \label{fig:models}
    \end{figure}

    \subsection{\ion{Ca}{ii} IRT variability}

    Figure~\ref{fig:models} shows the evolution of the IRT comparing SDSS data of 2003 with archival data of 2014, taken with the ESO VLT/\mbox{X-shooter} instrument \citep{Vernet:2011}. In the recent observation, the line profiles of all three IRT components show a narrow blue peak and a broad plateau on the red side. The former dominant red component now forms the plateau. The blue wing of the components is now very steep, whereas the red wing is broader than in the older observations.

\begin{figure}
      \centering
      \includegraphics[width=0.9\columnwidth]{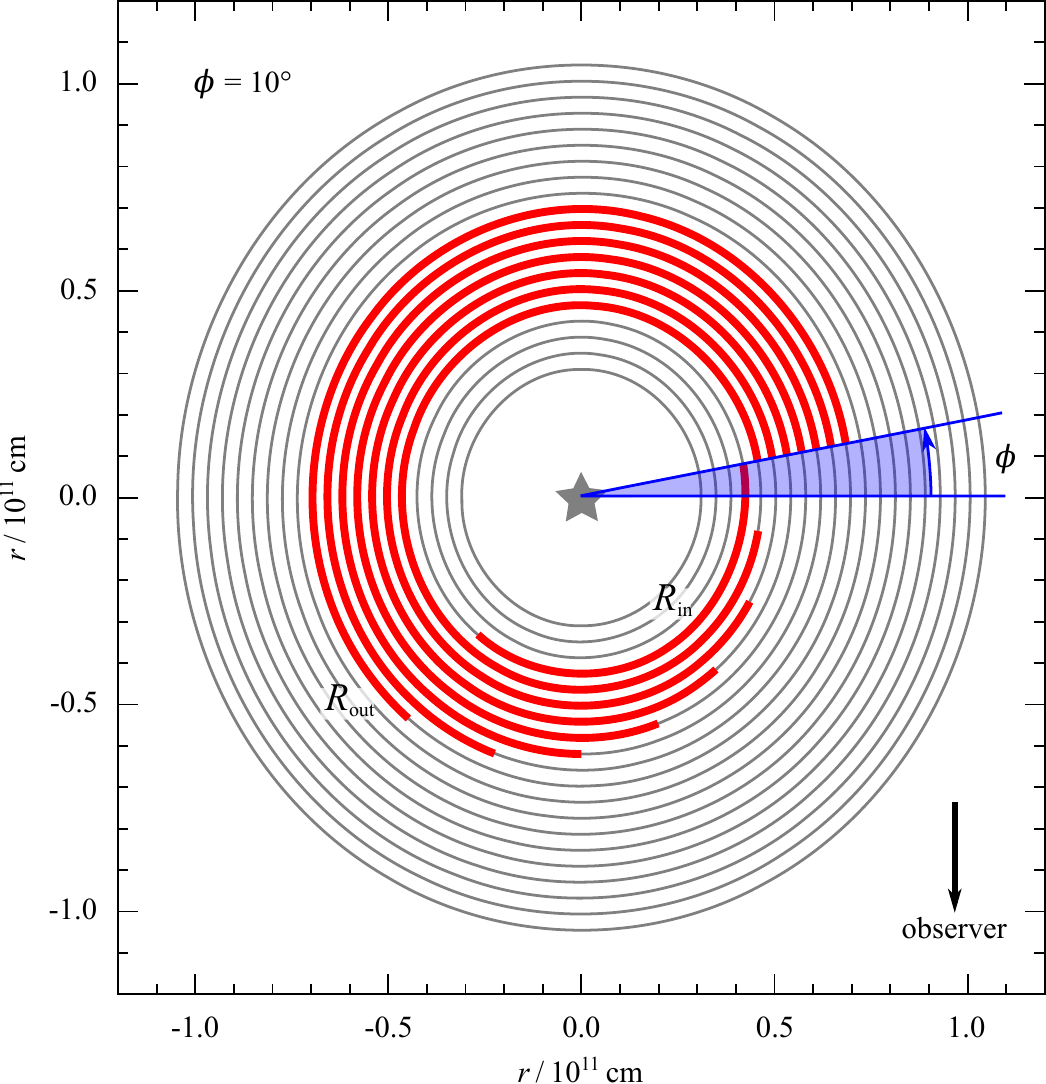}
      \caption{Schematics of the spiral-shape disk geometry. The viewing direction toward the observer is shown for $\phi=10^{\circ}$.}
      \label{fig:shape}
    \end{figure}

    To compare the observed spectra with our disk models we calculated a grid of different disk sizes and used the summed equivalent width of all three \ion{Ca}{ii} IRT components to determine the best fit. We assumed a bulk Earth-like mixture with reduced S for all models. Following our previous work \citep{Hartmann:2011,Hartmann:2014} we used an asymmetric disk geometry consisting of eight rings between $R_{\mathrm{in}}=4.24\times10^{10}$\,cm and $R_{\mathrm{out}}=6.93\times10^{10}$\,cm (ring numbers four to eleven) arranged in a spiral form as shown in Fig.\,\ref{fig:shape}. By altering the azimuthal viewing angle $\phi$ of the observer in steps of $\Delta\phi=5^{\circ}$, we found the best fit for the 2004 data to be $\phi=10^{\circ}$, and $\phi=120^{\circ}$ for the 2014 data, repectively. The overall change of the IRT line profile is reproduced quite well, nevertheless the blue peaks in the model spectra for the 2003 observation are slightly too narrow, whereas for the 2014 data the red peaks of the model are not as broad as those in the observation. On the other hand, the sharp blue wing of the 2014 spectrum is reproduced by the model.

    \citet{Manser:2016} discovered that some disk emission-lines show an asymmetry opposite to the \ion{Ca}{ii} IRT, indicating that the intensity distribution in the disk is not the same for each ion. One example is the \ion{O}{i} $\lambda$\,8446\,\AA\ line that in their Fig.\,3 displays a stronger red component while the \ion{Ca}{ii} IRT has a dominant blue component. As an experiment we have modified our spiral-shape model (Fig.\,\ref{fig:shape}) by augmenting the "empty" parts of the rings \mbox{4--11} with cooler regions ($\Teff=5000$\,K, same surface density as the 6000\,K regions). This preliminary model (not presented in this work) indeed shows the observed opposite line symmetries, because now the \ion{Ca}{ii} IRT asymmetry switches. The reason is that the cooler disk regions emit stronger in the \ion{Ca}{ii} IRT than the warmer regions.

    Assuming a homogeneous precession of the disk, we derive from the comparison of our models with the 2003 and 2014 spectra a period of $t_{\phi}=37\pm 3\,$a. This is similar to the precession period of 24--30\,a found by \citet{Manser:2016}.

    Based on our model, we can estimate the mass of the gas disk. The surface area of the disk, with the asymmetric geometry shown in Fig.\,\ref{fig:shape}, is $7.7\times 10^{21}$\,cm$^2$. With $\varSigma=0.3$\,g\,cm$^{-2}$, the mass of the gaseous disk is $2.3 \times 10^{21}$\,g, about 10\% of the estimated dusty debris disk's mass \citep{Brinkworth:2009}. It is possible that this gas mass may change if, for example, the \ion{Z}{ii} modeling approach was used.

    \section{Summary and conclusion}\label{sec:summary}

    Our search for disk emission-lines from \sdss\ in the FUV wavelength range was unsuccessful. This suggests that the disk is cooler than previously assumed, namely $\Teff\approx 5000$\,K and not $\approx 6000$\,K.

    From the analysis of their UV spectra (performed at longer wavelengths than our observation), \citet{Gaensicke:2012} conclude that the material accreted by the WD resembles a bulk Earth-like mixture. This is confirmed by our models otherwise, for a chondritic mixture, carbon emission lines should be visible in the vicinity of the \ion{Ca}{ii} IRT. The same holds for sulfur: The \mbox{non-detection} of S lines suggest an underabundance compared to bulk Earth value, so that the abundance of this volatile element resembles more the Earth mantle \citep{Allegre:1995}. The element is not detected in the WD photosphere but this does not allow to significantly constrain the sulfur abundance in the accreted material \citep{Gaensicke:2012}.

    Applying our models with a spiral shape to explain the time evolution of the \ion{Ca}{ii} IRT between 2003 and 2014 suggests a precession of the spiral pattern with a period of $\approx 37$\,a, consistent with the result of \citet{Manser:2016}. Our models also confirm their suggestion that the observed different red/blue asymmetries of disk line profiles can be explained by a \mbox{non-axisymmetric} temperature distribution.

    \begin{acknowledgements}
      S.H. was supported by the German Research Foundation (DFG, grant We 1312/37-1), T.R. by the German Aerospace Center (DLR, grant 05\,OR\,1402). The TMAD service (\url{http://astro-uni-tuebingen.de/~TMAD}) used to compile atomic data for this paper was constructed as part of the activities of the German Astrophysical Virtual Observatory. Some of the data presented in this paper were obtained from the Mikulski Archive for Space Telescopes (MAST). STScI is operated by the Association of Universities for Research in Astronomy, Inc., under NASA contract NAS5-26555. Support for MAST for non-HST data is provided by the NASA Office of Space Science via grant NNX09AF08G and by other grants and contracts. Based on observations made with ESO Telescopes at the La Silla Paranal Observatory under programme ID 093.D-0426. Funding for SDSS-III has been provided by the Alfred P. Sloan Foundation, the Participating Institutions, the National Science Foundation, and the U.S. Department of Energy Office of Science. The SDSS-III web site is http://www.sdss3.org/. This research made use of the SIMBAD database, operated at CDS, Strasbourg, France, and of NASA's Astrophysics Data System.

    \end{acknowledgements}

    \bibliographystyle{aa}
    \bibliography{hartmann}

\begin{thebibliography}{29}
\expandafter\ifx\csname natexlab\endcsname\relax\def\natexlab#1{#1}\fi

\bibitem[{{All{\`e}gre} {et~al.}(1995){All{\`e}gre}, {Poirier}, {Humler}, \&
  {Hofmann}}]{Allegre:1995}
{All{\`e}gre}, C.~J., {Poirier}, J.-P., {Humler}, E., \& {Hofmann}, A.~W. 1995,
  Earth and Planetary Science Letters, 134, 515

\bibitem[{{Brinkworth} {et~al.}(2009){Brinkworth}, {G{\"a}nsicke}, {Marsh},
  {Hoard}, \& {Tappert}}]{Brinkworth:2009}
{Brinkworth}, C.~S., {G{\"a}nsicke}, B.~T., {Marsh}, T.~R., {Hoard}, D.~W., \&
  {Tappert}, C. 2009, \apj, 696, 1402

\bibitem[{{Debes} \& {Sigurdsson}(2002)}]{Debes:2002}
{Debes}, J.~H. \& {Sigurdsson}, S. 2002, \apj, 572, 556

\bibitem[{{Farihi} {et~al.}(2012){Farihi}, {G{\"a}nsicke}, {Steele}, {Girven},
  {Burleigh}, {Breedt}, \& {Koester}}]{Farihi:2012}
{Farihi}, J., {G{\"a}nsicke}, B.~T., {Steele}, P.~R., {et~al.} 2012, \mnras,
  421, 1635

\bibitem[{{Farihi} {et~al.}(2010){Farihi}, {Jura}, {Lee}, \&
  {Zuckerman}}]{Farihi:2010}
{Farihi}, J., {Jura}, M., {Lee}, J., \& {Zuckerman}, B. 2010, \apj, 714, 1386

\bibitem[{{G{\"a}nsicke} {et~al.}(2012){G{\"a}nsicke}, {Koester}, {Farihi},
  {Girven}, {Parsons}, \& {Breedt}}]{Gaensicke:2012}
{G{\"a}nsicke}, B.~T., {Koester}, D., {Farihi}, J., {et~al.} 2012, \mnras, 424,
  333

\bibitem[{{G{\"a}nsicke} {et~al.}(2008){G{\"a}nsicke}, {Koester}, {Marsh},
  {Rebassa-Mansergas}, \& {Southworth}}]{Gaensicke:2008}
{G{\"a}nsicke}, B.~T., {Koester}, D., {Marsh}, T.~R., {Rebassa-Mansergas}, A.,
  \& {Southworth}, J. 2008, \mnras, 391, L103

\bibitem[{{G{\"a}nsicke} {et~al.}(2006){G{\"a}nsicke}, {Marsh}, {Southworth},
  \& {Rebassa-Mansergas}}]{Gaensicke:2006}
{G{\"a}nsicke}, B.~T., {Marsh}, T.~R., {Southworth}, J., \&
  {Rebassa-Mansergas}, A. 2006, Science, 314, 1908

\bibitem[{{Hartmann} {et~al.}(2011){Hartmann}, {Nagel}, {Rauch}, \&
  {Werner}}]{Hartmann:2011}
{Hartmann}, S., {Nagel}, T., {Rauch}, T., \& {Werner}, K. 2011, \aap, 530, A7

\bibitem[{{Hartmann} {et~al.}(2014){Hartmann}, {Nagel}, {Rauch}, \&
  {Werner}}]{Hartmann:2014}
{Hartmann}, S., {Nagel}, T., {Rauch}, T., \& {Werner}, K. 2014, \aap, 571, A44

\bibitem[{{Jura}(2003)}]{Jura:2003}
{Jura}, M. 2003, \apjl, 584, L91

\bibitem[{{Jura}(2008)}]{Jura:2008}
{Jura}, M. 2008, \aj, 135, 1785

\bibitem[{{Jura} \& {Young}(2014)}]{Jura:2014}
{Jura}, M. \& {Young}, E.~D. 2014, Annual Review of Earth and Planetary
  Sciences, 42, 45

\bibitem[{{Koester}(2009)}]{Koester:2009}
{Koester}, D. 2009, \aap, 498, 517

\bibitem[{{Koester} {et~al.}(2014){Koester}, {G{\"a}nsicke}, \&
  {Farihi}}]{Koester:2014}
{Koester}, D., {G{\"a}nsicke}, B.~T., \& {Farihi}, J. 2014, \aap, 566, A34

\bibitem[{{Koester} {et~al.}(2005){Koester}, {Napiwotzki}, {Voss}, {Homeier},
  \& {Reimers}}]{Koester:2005}
{Koester}, D., {Napiwotzki}, R., {Voss}, B., {Homeier}, D., \& {Reimers}, D.
  2005, \aap, 439, 317

\bibitem[{{Kromer} {et~al.}(2007){Kromer}, {Nagel}, \& {Werner}}]{Kromer:2007}
{Kromer}, M., {Nagel}, T., \& {Werner}, K. 2007, \aap, 475, 301

\bibitem[{{Manser} {et~al.}(2016){Manser}, {G{\"a}nsicke}, {Marsh}, {Veras},
  {Koester}, {Breedt}, {Pala}, {Parsons}, \& {Southworth}}]{Manser:2016}
{Manser}, C.~J., {G{\"a}nsicke}, B.~T., {Marsh}, T.~R., {et~al.} 2016, \mnras,
  455, 4467

\bibitem[{{Melis} {et~al.}(2010){Melis}, {Jura}, {Albert}, {Klein}, \&
  {Zuckerman}}]{Melis:2010}
{Melis}, C., {Jura}, M., {Albert}, L., {Klein}, B., \& {Zuckerman}, B. 2010,
  \apj, 722, 1078

\bibitem[{{Nagel} {et~al.}(2004){Nagel}, {Dreizler}, {Rauch}, \&
  {Werner}}]{Nagel:2004}
{Nagel}, T., {Dreizler}, S., {Rauch}, T., \& {Werner}, K. 2004, \aap, 428, 109

\bibitem[{{Rauch} \& {Deetjen}(2003)}]{Rauch:2003}
{Rauch}, T. \& {Deetjen}, J.~L. 2003, in Astronomical Society of the Pacific
  Conference Series, Vol. 288, Stellar Atmosphere Modeling, ed. I.~{Hubeny},
  D.~{Mihalas}, \& K.~{Werner}, 103

\bibitem[{{Shakura} \& {Sunyaev}(1973)}]{Shakura:1973}
{Shakura}, N.~I. \& {Sunyaev}, R.~A. 1973, \aap, 24, 337

\bibitem[{{Vernet} {et~al.}(2011){Vernet}, {Dekker}, {D'Odorico}, {Kaper},
  {Kjaergaard}, {Hammer}, {Randich}, {Zerbi}, {Groot}, {Hjorth}, {Guinouard},
  {Navarro}, {Adolfse}, {Albers}, {Amans}, {Andersen}, {Andersen}, {Binetruy},
  {Bristow}, {Castillo}, {Chemla}, {Christensen}, {Conconi}, {Conzelmann},
  {Dam}, {de Caprio}, {de Ugarte Postigo}, {Delabre}, {di Marcantonio},
  {Downing}, {Elswijk}, {Finger}, {Fischer}, {Flores}, {Fran{\c c}ois},
  {Goldoni}, {Guglielmi}, {Haigron}, {Hanenburg}, {Hendriks}, {Horrobin},
  {Horville}, {Jessen}, {Kerber}, {Kern}, {Kiekebusch}, {Kleszcz}, {Klougart},
  {Kragt}, {Larsen}, {Lizon}, {Lucuix}, {Mainieri}, {Manuputy}, {Martayan},
  {Mason}, {Mazzoleni}, {Michaelsen}, {Modigliani}, {Moehler}, {M{\o}ller},
  {Norup S{\o}rensen}, {N{\o}rregaard}, {P{\'e}roux}, {Patat}, {Pena}, {Pragt},
  {Reinero}, {Rigal}, {Riva}, {Roelfsema}, {Royer}, {Sacco}, {Santin},
  {Schoenmaker}, {Spano}, {Sweers}, {Ter Horst}, {Tintori}, {Tromp}, {van
  Dael}, {van der Vliet}, {Venema}, {Vidali}, {Vinther}, {Vola}, {Winters},
  {Wistisen}, {Wulterkens}, \& {Zacchei}}]{Vernet:2011}
{Vernet}, J., {Dekker}, H., {D'Odorico}, S., {et~al.} 2011, \aap, 536, A105

\bibitem[{{Werner}(1986)}]{Werner:1986}
{Werner}, K. 1986, \aap, 161, 177

\bibitem[{{Werner} {et~al.}(2003){Werner}, {Deetjen}, {Dreizler}, {Nagel},
  {Rauch}, \& {Schuh}}]{Werner:2003}
{Werner}, K., {Deetjen}, J.~L., {Dreizler}, S., {et~al.} 2003, in Astronomical
  Society of the Pacific Conference Series, Vol. 288, Stellar Atmosphere
  Modeling, ed. I.~{Hubeny}, D.~{Mihalas}, \& K.~{Werner}, 31

\bibitem[{{Werner} {et~al.}(2012){Werner}, {Dreizler}, \&
  {Rauch}}]{Werner:2012}
{Werner}, K., {Dreizler}, S., \& {Rauch}, T. 2012, {TMAP: T{\"u}bingen NLTE
  Model-Atmosphere Package}, Astrophysics Source Code Library

\bibitem[{{Werner} {et~al.}(2007){Werner}, {Nagel}, \& {Rauch}}]{Werner:2007}
{Werner}, K., {Nagel}, T., \& {Rauch}, T. 2007, \apss, 308, 141

\bibitem[{{Werner} {et~al.}(2006){Werner}, {Nagel}, {Rauch}, {Hammer}, \&
  {Dreizler}}]{Werner:2006}
{Werner}, K., {Nagel}, T., {Rauch}, T., {Hammer}, N.~J., \& {Dreizler}, S.
  2006, \aap, 450, 725

\bibitem[{{Wilson} {et~al.}(2014){Wilson}, {G{\"a}nsicke}, {Koester}, {Raddi},
  {Breedt}, {Southworth}, \& {Parsons}}]{Wilson:2014}
{Wilson}, D.~J., {G{\"a}nsicke}, B.~T., {Koester}, D., {et~al.} 2014, \mnras,
  445, 1878

\end{thebibliography}

\end{document}